\documentclass[aps,prl,twocolumn,showpacs]{revtex4}
\usepackage{txfonts}
\usepackage{graphicx}

\begin{document}

\title{Fully Gapped Superconducting State Based on a High Normal State Quasiparticle Density of States in Ba$_{0.6}$K$_{0.4}$Fe$_2$As$_2$ Single Crystals}

\author{Gang Mu, Huiqian Luo, Zhaosheng Wang, Lei Shan, Cong Ren and Hai-Hu Wen}\email{hhwen@aphy.iphy.ac.cn }

\affiliation{National Laboratory for Superconductivity, Institute of
Physics and Beijing National Laboratory for Condensed Matter
Physics, Chinese Academy of Sciences, P.O. Box 603, Beijing 100190,
People's Republic of China}

\begin{abstract}
We report the specific heat (SH) measurements on single crystals of
hole doped $FeAs$-based superconductor $Ba_{0.6}K_{0.4}Fe_2As_2$. It
is found that the electronic SH coefficient $\gamma_e(T)$ is not
temperature dependent and increases almost linearly with the
magnetic field in low temperature region. These point to a fully
gapped superconducting state. Surprisingly the sharp SH anomaly
$\Delta C/T|_{T_c}$ reaches a value of 98 $mJ/mol K^2$ suggesting a
very high normal state quasiparticle density of states ($\gamma_n
\approx 63 mJ/mol K^2$ ). A detailed analysis reveals that the
$\gamma_e(T)$ cannot be fitted with a single gap of s-wave symmetry
due to the presence of a hump in the middle temperature region.
However, our data indicate that the dominant part of the
superconducting condensate is induced by an s-wave gap with the
magnitude of about 6 meV.

\end{abstract} 
\maketitle

The discovery of high temperature superconductivity in the
$FeAs$-based system has stimulated enormous interests in the field
of condensed matter physics and material sciences
\cite{Kamihara2008}. The superconductivity has not only been
discovered in the electron doped samples, but also in the hole-doped
ones\cite{Wen2008,Rotter1}. The central issues concerning the
superconductivity mechanism are about the symmetry and the magnitude
of the superconducting gap. The experimental results obtained so far
are, however, highly controversial. The low temperature specific
heat (SH) measurements in the F-doped $LaFeAsO$ samples revealed a
nonlinear magnetic field dependence of the SH coefficient
$\gamma_e$, which was attributed to the presence of a nodal
gap\cite{MuG}. This was later supported by many other measurements
based on $\mu$SR\cite{uSR1,uSR2,uSR3}, NMR\cite{NMR1,NMR2,NMR3},
magnetic penetration\cite{Hc1} and point contact Andreev spectrum
(PCAS)\cite{ShanL}. On the other hand, the PCAS on the F-doped
SmFeAsO indicated a feature of s-wave gap\cite{Chien}, some
measurements\cite{swave1,swave2,swave3,swave4} also gave support to
this conclusion. It is important to note that most of the
conclusions drawn for a nodal gap were obtained on the electron
doped LnFeAsO samples (abbreviated as $FeAs$-1111, Ln stands for the
rare earth elements) which are characterized by a low charge carrier
density and thus low superfluid density\cite{ZhuXY}. For the
$FeAs$-1111 phase, it is very difficult to grow crystals with large
sizes, therefore most of the measurements on the pairing symmetry so
far were made on polycrystalline samples. This is much improved in
the $(Ba,Sr)_{1-x}K_xFe_2As_2$ (denoted as $FeAs$-122) system since
sizable crystals can be achieved\cite{Ni,LuoHQ}. Preliminary data by
angle resolved photo-emission spectroscopy (ARPES) on these crystals
show two groups of superconducting gaps ($\Delta_1\approx$ 12 meV,
$\Delta_2 \approx$ 6 meV) all with s-wave
symmetry\cite{DingH,ZhouXJ,Hasan}. It is known that the surface of
this type of single crystals decay or reconstruct very quickly, this
may give obstacles to get repeatable data when using the surface
sensitive tools. Thus solid conclusions about the gap symmetry and
magnitude from bulk measurements are highly desired.

Specific heat (SH) is one of the powerful tools to measure the
quasiparticle density of states (DOS) at the Fermi level. By
measuring the variation of the electronic SH versus temperature and
magnetic field, one can essentially determine the feature of the gap
symmetry. In this Letter, for the first time, we report the detailed
low temperature SH data on $Ba_{0.6}K_{0.4}Fe_2As_2$ single crystals
with $T_c$ = 36.5 K (90\%$\rho_n$). Our results elucidate a fully
gaped feature of the superconducting state. Meanwhile we show the
evidence of a large DOS for the normal state of the $FeAs$-122
superconductors, which is in sharp contrast with that of the
$FeAs$-1111 phase.

The superconducting single crystals with $T_c$ of about 36.5 K were
grown by the self-flux method\cite{LuoHQ}. The sample for the SH
measurement has the dimensions of 3 $\times$ 1.5 $\times$ 0.2
mm$^{3}$. The resistivity and the specific heat were measured with a
Quantum Design instrument physical property measurement system
(PPMS) with the temperature down to 1.8 K and the magnetic field up
to 9 T. We employed the thermal relaxation technique to perform the
specific heat measurements. To improve the resolution, we used a
latest developed SH measuring puck from Quantum Design, which has
negligible field dependence of the sensor of the thermometer on the
chip as well as the thermal conductance of the thermal linking
wires.

In the main panel of Fig. 1 we show the raw data of SH coefficient
$\gamma=C/T$ vs $T$ at 0 T and 9 T. Multiple complicated
contributions to the SH data emerged in the low-T region when a
magnetic field was applied, so we only showed the data above 4.3 K
under magnetic fields, and the data at zero field was shown down to
about 1.8 K. Clear and sharp superconducting anomalies can be seen
near $T_c$ from the raw data. The SH anomaly $\Delta C/T|_{T_c}$ at
zero field was determined to be about 98 mJ/mol K$^2$, indicated by
the vertical short blue line in the inset (a). This is remarkably
different from the case in the $FeAs$-1111 phase, where no visible
or only very small SH anomaly were observed in the raw
data\cite{MuG,David,SYLi}. Even in the same $FeAs$-122 system, the
magnitude of the anomaly in our data is several times larger than
that observed in the polycrystalline samples\cite{SYLi2} and the
single crystals grown using Sn as the flux\cite{Ni}. This large
value of $\Delta C/T|_{T_c}$ clearly suggests a rather high normal
state quasiparticle DOS in this system, which will be further
addressed later. A magnetic field of 9 T shifts the SH anomaly down
for only 1.5 K and suppresses the anomaly. In the low-T region, a
clear flattening feature of $C/T$ can be seen in the zero field
data, which may imply the weak excitation of quasiparticles. The
temperature dependence of the resistivity at different magnetic
fields are shown in the inset (b) of Fig. 1. By applying a magnetic
field the middle transition point (50\%$\rho_n$) shifts to lower
temperatures slowly with a slope $-d \mu_0H_{c2}(T)/dT|_{T_c}
\approx$ 4.1 T / K. Using the Werthamer-Helfand-Hohenberg
relation\cite{WHH} $\mu_0H_{c2}(0)=-0.69d \mu_0H_{c2}(T)/dT|_{T_c}
T_c$, we get the upper critical field $\mu_0H_{c2}(0) \approx$ 100 T
($H\|c$).

\begin{figure}
\includegraphics[width=8cm]{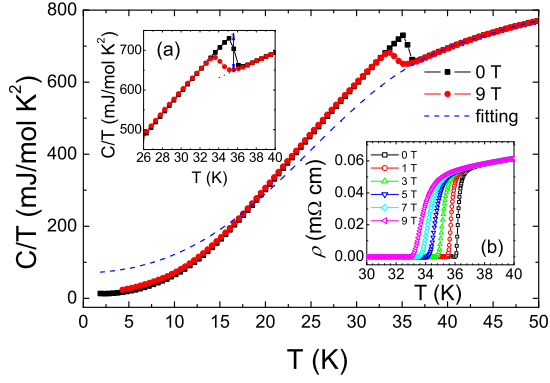}
\caption {(color online) Raw data of SH coefficient $\gamma=C/T$ vs
$T$ are shown in the main frame. The dashed line shows the normal
state SH obtained from fitting to eq.(4). The inset (a) shows an
enlarged view of the data $\gamma=C/T$ near $T_c$. The sharp SH
anomaly $\Delta C/T|_{T_c}$ is indicated by the arrowed blue short
line with a magnitude of about 98 mJ/mol K$^2$. The inset (b)
presents the resistive transition curves at magnetic fields ranging
from 0 to 9 T. } \label{fig1}
\end{figure}

\begin{figure}
\includegraphics[width=8cm]{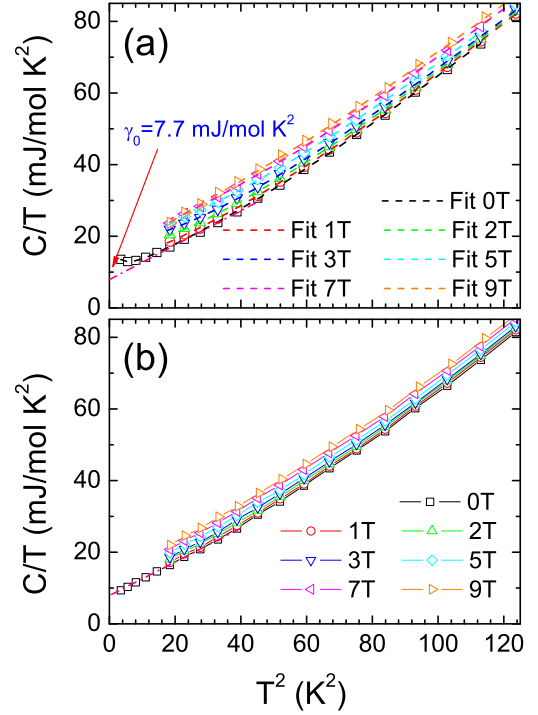}
\caption {(color online) Temperature and magnetic field dependence
of specific heat in $C/T$ vs $T^2$ plot in the low-T range. (a) Raw
data before removing the Schottky anomaly. The dashed lines
represent the theoretical fit (see text) containing all terms in Eq.
(2). (b) Replot of the data after the Schottky anomaly was
subtracted. The dot-dashed line represents a extension of the zero
field data to T = 0 K giving a residual value $\gamma_0$ = 7.7
mJ/mol K$^2$(see text).} \label{fig2}
\end{figure}

The raw data in the low-T region at different fields are plotted as
$C/T$ vs $T^2$ in Fig. 2(a). Detailed analysis reveals that weak
Schottky anomaly still contributes to the whole SH in the low-T
region. A slight curvature was also detected in the region from 4.3
K to 11 K on the plot of $C/T$ vs. $T^2$, which was attributed to
the electron contribution of the superconducting state and the
quintic term of the phonon contribution. This brought in enormous
difficulties when treating the data because it gave too many fitting
parameters. Therefore we first analyze the data below 6 K at zero
field, where the two terms mentioned above remain negligible.
Consequently the zero field data below 6 K can be represented by the
following equation:
\begin{equation}
C(T,H=0)=\gamma_0 T+\beta T^3+C_{Sch}(T,H=0),\label{eq:1}
\end{equation}
where the three terms represent the contributions of the residual
electronic SH, the phonon and the magnetic impurity (the so-called
Schottky anomaly), respectively. The two-level Schottky anomaly is
given by $nx^2e^x/(1+e^x)^2$ $(x=g\mu_B H_{eff}/k_B T)$, where g is
the Land\'{e} factor, $\mu_B$ is the Bohr magneton,
H$_{eff}$=$\sqrt{H^2+H_{0}^{2}}$ is the effective magnetic field
which evolves into $H_{eff}$ = $H_0$, the crystal field at zero
external field, and $n$ is the concentration of paramagnetic
centers. The obtained fitting parameters $\gamma_0 \approx$ 7.7
mJ/mol K$^2$ and $\beta\approx$ 0.473 mJ/mol K$^4$ are very close to
the values obtained by simply drawing a linear line below 6 K as
shown by the dot-dashed line in Fig. 2(a). The values of $\gamma_0$
and $\beta$ are then fixed when fitting the zero field data up to 11
K, where all the terms must be taken into account:
\begin{equation}
C(T,H=0)=\gamma_0 T+[\beta T^3+\eta
T^5]+C_{es}+C_{Sch}(T,H=0),\label{eq:2}
\end{equation}
where $\eta$ is the quintic term coefficient of the phonon SH and
$C_{es}=D\times e^{-\Delta(0)/k_B T}/T^{1.5}$ is the superconducting
electron contribution, with $\Delta(0)$ the superconducting gap at 0
K. By fitting the data at zero field using Eq. (2), we obtained
$\eta \approx$ 0.00034 mJ/mol K$^6$ and $\Delta(0) \approx$ 5.99
$\pm$ 0.3 meV. As for the data under finite fields, a magnetic field
induced term $\gamma(H)$ arises and the total SH can be written as
\begin{equation}
C(T,H)=[\gamma_0+\gamma(H)] T+[\beta T^3+\eta
T^5]+C_{es}+C_{Sch}(T,H).\label{eq:3}
\end{equation}
It is quite rational to fix $\gamma_0$, $\beta$, and $\eta$ as the
values obtained from analyzing the data at zero field. And the
superconducting gap under magnetic fields was restricted using the
relation\cite{Maki} $\Delta(H)=\Delta(0)\sqrt{1-H/H_{c2}}$ assuming
a field induced pair breaking effect. In this way the number of the
fitting parameters were reduced remarkably and creditable results
can be obtained.

\begin{figure}
\includegraphics[width=8cm]{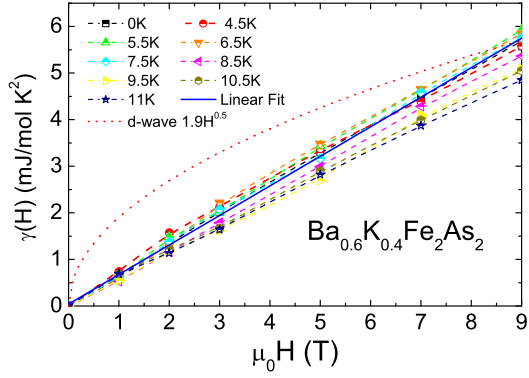}
\caption{(color online) Field dependence of the field-induced term
$\gamma(H)$ at temperatures ranging from 4.5 K to 11.1 K, and that
at $T = 0$ K obtained from fitting (see text). The dashed lines for
different temperatures are guides for the eyes. The blue solid line
is a linear fit to the zero temperature data, and the red dotted
line is a fit to the d-wave prediction $\gamma(H)$ = A$\sqrt{H}$. }
\label{fig3}
\end{figure}

Fig. 2(b) shows the data after removing the Schottky anomaly from
the total SH. The obtained field induced term $\gamma(H)$ are shown
in Fig. 3 (will be discussed later) and the fitting parameters
related to the terms $C_{es}$ and $C_{Sch}$ are shown in Table I.
The obtained residual term $\gamma_0 \approx$ 7.7 mJ/mol K$^2$
accounts for about 11\% of the total electron contribution (will be
discussed later), indicating a superconducting volume fraction of
about 89\% in our sample. Using the obtained value of $\beta$ and
the relation $\Theta_D$ = $(12\pi^4k_BN_AZ/5\beta)^{1/3}$, where
$N_A$ = 6.02 $\times 10^{23}$ mol$^{-1}$ is the Avogadro constant, Z
= 5 is the number of atoms in one unit cell, we get the Debye
temperature $\Theta_D \approx$ 274 K. This value is comparable to
that found in the LaFeAsO$_{0.9}$F$_{0.1-\delta}$ system\cite{MuG}.
\begin{table}
\caption{Fitting parameters ($\Delta$ is calculated through
$\Delta(H)=\Delta(0)\sqrt{1-H/H_{c2}}$, $H_0$ is fixed with the
value from the fitting to zero-field data).}
\begin{tabular}{cccccc}
\hline \hline
$\mu_0H$(T) & $D$(mJ$\,$K$^{0.5}/$mol) & $\Delta$(meV) & $n$(mJ$/$mol$\,$K) & $\mu_0H_0$(T)  & $g$ \\
\hline
0.0      & $2.11\times10^{6}$       & $5.99$      & $20.95$        & $1.70$      & $2.75$ \\
1.0      & $2.07\times10^{6}$       & $5.96$      & $31.08$        & $1.70$      & $3.18$ \\
2.0      & $1.93\times10^{6}$       & $5.93$      & $35.76$        & $1.70$      & $3.26$ \\
3.0      & $1.82\times10^{6}$       & $5.90$      & $36.22$        & $1.70$      & $3.38$ \\
5.0      & $1.63\times10^{6}$       & $5.84$      & $36.11$        & $1.70$      & $3.24$ \\
7.0      & $1.57\times10^{6}$       & $5.78$      & $36.71$        & $1.70$      & $3.11$ \\
9.0      & $1.43\times10^{6}$       & $5.72$      & $35.98$        & $1.70$      & $3.22$ \\
 \hline \hline
\end{tabular}
\label{tab:table1}
\end{table}

The field-induced change of the electron SH coefficient $\gamma(H)$
was investigated carefully. This term at zero temperature obtained
from fitting, along with the data at finite temperatures extracted
from Fig. 2(b), is plotted in Fig. 3. It can be seen clearly that
$\gamma(H)$ increases almost linearly with the magnetic field in the
temperature region up to 11 K. A linear fit with the slope of about
0.633 $mJ/mol K^2$ T to the zero temperature data is revealed by the
blue solid line in this figure. This linear behavior is actually
anticipated by the theoretical prediction for superconductors with a
full gap\cite{Hussey}, in which $\gamma(H)$ is mainly contributed by
the localized quasiparticle DOS within vortex cores. This is in
sharp contrast with the results in cuprates\cite{Moler,WenHH} and
the LaFeAsO$_{0.9}$F$_{0.1-\delta}$ system\cite{MuG} where a
$\gamma(H)\propto\sqrt{H}$ relation was observed and attributed to
the Doppler shift of the nodal quasiparticle spectrum. The curve
plotted using the relation $\gamma(H)$ = A$\sqrt{H}$ is also
presented in Fig. 3 by the red dotted line for comparison. It is
obvious that this curve fails to fit our data.

\begin{figure}
\includegraphics[width=8cm]{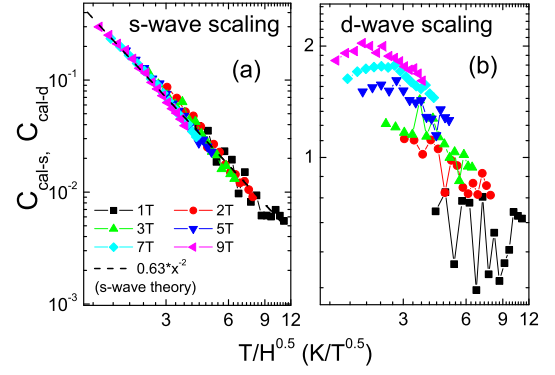}
\caption {(color online) (a) Scaling of the data according to the
s-wave scenario (symbols) $C_{cal-s}$ vs. T$/\sqrt{H}$, the dashed
line represents the theoretical expression. (b) Scaling of the data
(symbols) based on d-wave prediction $C_{cal-d}$ vs. T$/\sqrt{H}$.
No good scaling can be found for the d-wave case. } \label{fig4}
\end{figure}

In order to further confirm the gap symmetry, we analyzed the SH
data in finite temperature region in the mixed state. In s-wave
superconductors, the inner-core states dominate the quasiparticle
excitations, and consequently a simple scaling law
$C_{core}/T^3\approx (\gamma_n/H_{c2}(0))\times(T/\sqrt{H})^{-2}$ is
expected at low-T. While for a gap with line nodes, the excitation
spectrum is dominated by the extended quasiparticles outside the
vortex cores. And the so-called Simon-Lee scaling law\cite{SimonLee}
$C_{vol}/(T\sqrt{H})=f(T/\sqrt{H})$ may be obeyed. A simple analysis
similar to that has been done in our previous work\cite{LiuZY} and
shows that for the superconductor with an s-wave symmetry,
$C_{cal-s}=[(C(H)-C_{Sch}(H))-(C(H=0)-C_{Sch}(H=0))]/T^{3} \approx
C_{core}/T^{3}$. In other words, the defined term $C_{cal-s}$ should
scale with $(T/\sqrt{H})^{-2}$ with the prefactor
$\gamma_n/H_{c2}(0)$. Similarly for the d-wave symmetry we have
known\cite{LiuZY,WenHH} that
$C_{cal-d}=[(C(H)-C_{Sch}(H))-(C(H=0)-C_{Sch}(H=0))]/T\sqrt{H} =
C_{vol}/T\sqrt{H}-\alpha T/\sqrt{H}$, where $\alpha$ is the electron
SH coefficient at zero field for a d-wave superconductor, should
also scale with $T/\sqrt{H}$.

The scaling result with the s-wave condition is presented in Fig.
4(a). One can see that all the data at different magnetic fields can
be scaled roughly to one straight line, which reflects the
theoretical curve $C_{cal-s}= 0.633\times (T/\sqrt{H})^{-2}$.
Naturally, this prefactor $\gamma_n/H_{c2}(0)= 0.633$ mJ/mol K$^2$ T
is consistent with the magnitude of the slope of the blue line in
Fig. 3. Using the value of $H_{c2}(0) \approx 100$ T, we can
estimate the normal state electron SH coefficient $\gamma_n$ of
about 63.3 mJ/mol K$^2$. Fig.4(b) shows the scaling by following the
d-wave scheme. It is clear that the s-wave scaling is much better
than that of the d-wave case. So it seems that the superconducting
gap in this $FeAs$-122 phase has an s-wave symmetry and the
field-induced quasiparticle DOS are mainly contributed by the vortex
cores.

Using the value of $\gamma_n\approx$ 63.3 mJ/mol K$^2$, we get the
ratio $\Delta C_e/\gamma_nT|_{T_c} \approx $ 1.55 being very close
to the weak-coupling BCS value 1.43. Considering the electron
re-normalization effect, the electron SH coefficient of a metal can
be written as $\gamma_n=2\pi^2N(E_F)k_B^2(1+\lambda)/3$, where
$N(E_F)$ is the DOS at the Fermi surface and $\lambda$ reflects the
coupling strength. The fact that the electron-phonon coupling
strength is weak in present system indicates that the large value of
$\gamma_n$ is not due to the enhanced effective mass but originates
from the high normal state quasiparticle DOS. Comparing with the
$\gamma_n$ obtained for the F-doped LaFeAsO system\cite{MuG} (about
5-6 $mJ/mol-Fe K^2$), the $N(E_F)$ in hole doped $FeAs$-122 may be
3-5 times higher than that in the electron doped $FeAs$-1111. This
may give an important clue in the calculation of the $N(E_F)$
induced by the doping effect.
\begin{figure}
\includegraphics[width=8cm]{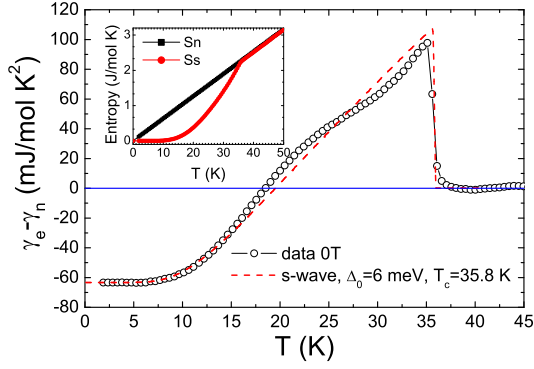}
\caption {(color online) Temperature dependence of the electronic SH
contribution (with the normal state part subtracted) is shown in the
main frame. A sharp SH anomaly can be seen here. A hump is clearly
seen in the middle temperature region. The red dashed line is a
theoretical curve based on the BCS expression with an s-wave gap of
6 meV. The inset shows the entropy of the superconducting state (red
circle symbols) and the normal state (dark square symbols). }
\label{fig5}
\end{figure}

In the $FeAs$-122 superconductors, it is challenging to measure the
normal state SH below $T_c$ due to the very high $H_{c2}$. In order
to have a comprehensive understanding to the normal state electronic
SH, we have attempted to fit the normal state SH above $T_c$ using a
polynomial function:
\begin{equation}
C_{n}=(\gamma_0 +\gamma_n) + \beta_3 T^3+\beta_5 T^5+\beta_7
T^7+\beta_9 T^9+\beta_{11} T^{11},\label{eq:4}
\end{equation}
where we took the values obtained already $\gamma_0 = 7.7$ mJ/mol
K$^2$, $\beta_3 = 0.473$ mJ/mol K$^4$, $\gamma_n$ = 63.3 mJ/mol
K$^2$. Other fitting parameters, $\beta_5$, $\beta_7$, $\beta_9$,
and $\beta_{11}$, were left free in the fitting process, yielding
the values of 3.72$\times$ 10$^{-4}$ mJ/mol K$^6$, -5.32$\times$
10$^{-7}$ mJ/mol K$^8$, 2.13$\times$ 10$^{-10}$ mJ/mol K$^{10}$, and
-2.90$\times$ 10$^{-14}$ mJ/mol K$^{12}$, respectively. It's worth
to note that the value of $\beta_5$ is very close to the value of
$\eta$ obtained before. The fitting result of the normal state SH is
displayed by the blue dashed line in the main frame of Fig. 1. The
data after subtracting the normal state SH is presented in the main
frame of Fig. 5. It was found that the entropy-conserving law was
satisfied naturally confirming the validity of our fitting, as shown
in the inset of Fig. 5. A clear flattening of $\gamma_e-\gamma_n$ in
the temperature region up to 7 K is observed indicating a fully
gapped superconducting state. Moreover, a hump is clearly seen in
the middle temperature region. We attempted to fit the data using
the BCS formula:
\begin{eqnarray}
\gamma_\mathrm{e}=\frac{4N(0)}{k_BT^{3}}\int_{0}^{\hbar\omega_D}\int_0^{2\pi}\frac{e^{\zeta/k_BT}}{(1+e^{\zeta/k_BT})^{2}}(\varepsilon^{2}+\nonumber\\
\nonumber\\
\Delta^{2}(\theta,T)-\frac{T}{2}\frac{d\Delta^{2}(\theta,T)}{dT})\,d\theta\,d\varepsilon,
\end{eqnarray}
where $\zeta=\sqrt{\varepsilon^2+\Delta^2(T,\theta)}$ and
$\Delta(T,\theta)=\Delta_0(T)$ for the s-wave symmetry. The red
dashed line in the main frame presented the fitting result. One can
see that the fitting curve with a gap value of about 6 meV matched
our data below 13 K perfectly, but failed to describe the hump
feature in the middle temperature region. This hump may be
attributed to the multi-gap effect which seems to appear in the
$FeAs$ based superconductors. But at this moment we can't exclude
the possibility that the hump is induced by the limited uncertainty
in getting the normal state phonon contribution. Nevertheless, the
fine fitting in wide temperature region strongly suggests that the
dominant part of the superconducting condensate is induced by an
s-wave gap with the magnitude of about 6 meV. Our results here seem
to be consistent with the ARPES data, both in symmetry and the small
gap\cite{DingH,ZhouXJ,Hasan}. But we have not found a large gap of
12 meV. This discrepancy may be induced by the different ways in
determining the gap. Future works are certainly required to
reconcile all these distinct results.

In summary, the low temperature specific heat measurements reveal
that the $Ba_{0.6}K_{0.4}Fe_2As_2$ superconductor has an s-wave
pairing symmetry with the gap amplitude of about 6 meV. The rather
high SH anomaly $\Delta C/T|_{T_c}$ and the large value of
$\gamma_n$ suggest a high normal state quasiparticle DOS in this
system. This makes it very different from the $FeAs$-1111 phase. A
multigap feature seems possible but the other gap should be
responsible for only a small fraction of the superfluid density.

\begin{acknowledgments}
We acknowledge the fruitful discussions with Tao Xiang, Fuchun
Zhang, Huen-Dung Yang and Dung-Hai Lee. This work is supported by
the Natural Science Foundation of China, the Ministry of Science and
Technology of China (973 project No: 2006CB601000, 2006CB921107,
2006CB921802), and Chinese Academy of Sciences (Project ITSNEM).
\end{acknowledgments}


\begin{thebibliography}{00}

\bibitem{Kamihara2008}Y. Kamihara et al., JACS {\bf130}, 3296 (2008).
\bibitem{Wen2008}H. H. Wen et al., EPL {\bf82}, 17009 (2008).
\bibitem{Rotter1}M. Rotter et al., arXiv. Condat/0805. 4630. and arXiv. Condat/0807. 4096.
\bibitem{MuG}G. Mu et al., Chin. Phys. Lett. {\bf25}, 2221 (2008).
\bibitem{uSR1}H. Luetkens et al., arXiv: cond-mat/0804. 3115.
\bibitem{uSR2} J. P. Carlo, et al., arXiv: cond-mat/0805.2186.
\bibitem{uSR3}A. J. Drew et al., arXiv: cond-mat/0805.2186.
\bibitem{NMR1}H.-J. Grafe et al., arXiv: cond-mat/0805.2595.
\bibitem{NMR2} K. Matano et al., arXiv: cond-mat/0806.0249.
\bibitem{NMR3}K. Ahilan et al., arXiv: cond-mat/0804.4026.
\bibitem{Hc1}C. Ren et al., arXiv: cond-mat/0804.1726.
\bibitem{ShanL}L. Shan et al., arXiv: cond-mat/0803.2405.
\bibitem{Chien}T. Y. Chen et al., Nature {\bf 453}, 1224 (2008).
\bibitem{swave1}K. Hashimoto et al., arXiv: cond-mat/0806.3149.
\bibitem{swave2}A. A. Aczel et al., arXiv: cond-mat/0807.1044.
\bibitem{swave3}C. Martin arXiv: cond-mat/0807.0876.
\bibitem{swave4}L. Malone et al., arXiv: cond-mat/0806.3908.
\bibitem{ZhuXY}X. Y. Zhu et al., Supercond. Sci. Tech. 21, 105001
(2008).
\bibitem{Ni}N. Ni et al., Phys. Rev. B{\bf  78}, 014507 (2008)
\bibitem{LuoHQ}H. Q. Luo et al., arXiv:cond-mat/0807.0759.
\bibitem{DingH}H. Ding et al., EPL {\bf 83}, 47001
(2008).
\bibitem{ZhouXJ}L. Zhao et al., arXiv: cond-mat/0807.0398.
\bibitem{Hasan}L. Wray et al., arXiv: cond-mat/0808.2185.

\bibitem{David} Athena S. Sefat et al., arXiv:cond-mat/0803.2528.
\bibitem{SYLi} L. Ding et al., arXiv:cond-mat/0804.3642.
\bibitem{SYLi2} J. K. Dong et al., arXiv:cond-mat/0806.3573.
\bibitem{WHH}N. R. Werthamer, E. Helfand and P. C. Hohenberg, Phys. Rev. \textbf{147}, 295 (1966).
\bibitem{Maki}K. Maki, \emph{in Superconductivity}, edited by R. D. Parks (Marcel
Dekker, New York, 1969), Vol. 2, p. 1035.
\bibitem{Hussey}N. E. Hussey, Adv. Phys. {\bf51}, 1685 (2002)
\bibitem{Moler}K. A. Moler et al., Phys. Rev. Lett.\textbf{73}, 2744 (1994). K. A. Moler et al., Phys. Rev. B \textbf{55}, 12753 (1997).
\bibitem{WenHH}H. H. Wen et al., Phys. Rev. B \textbf{70}, 214505 (2004). H. H. Wen et al., Phys. Rev. B \textbf{72}, 134507 (2005).
\bibitem{SimonLee}S. H. Simon and P. A. Lee, Phys. Rev. Lett. \textbf{78}, 1548 (1997).
\bibitem{LiuZY}Z. Y. Liu et al., Europhys. Lett. {\bf69}, 263 (2005).





\end{thebibliography}
\end{document}